\documentclass[twocolumn]{aastex63}
\usepackage{CJK}
\usepackage{epstopdf}
\usepackage{graphicx}
\usepackage{subfigure}
\usepackage{chngcntr}

\newcommand{\y}[1]{{\textcolor{red}{#1}}}

\begin{document}

\title{On the Migration Origin of the Hercules Moving Group with GAIA, LAMOST, APOGEE, and GALAH Surveys}

\author[0000-0001-9283-8334]{Xilong Liang}
\altaffiliation{Department of Astronomy \& Center for Galaxy Evolution Research, Yonsei University, Seoul 03722, Republic of Korea}
\altaffiliation{National Astronomical Observatories, Chinese Academy of Sciences, Beijing 100101, China}

\author[0000-0002-1842-4325]{Suk-Jin Yoon}
\altaffiliation{Department of Astronomy \& Center for Galaxy Evolution Research, Yonsei University, Seoul 03722, Republic of Korea, sjyoon0691@yonsei.ac.kr}

\author[0000-0003-2868-8276]{Jingkun Zhao}
\altaffiliation{Key Laboratory of Optical Astronomy, National Astronomical Observatories, Chinese Academy of Sciences, Beijing 100012, China}

\author[0000-0001-5017-7021]{Zhaoyu Li}
\altaffiliation{Department of Astronomy, School of Physics and Astronomy, Shanghai Jiao Tong University, 800 Dongchuan Road, Shanghai 200240, China}

\author[0000-0003-1352-7226]{Jiajun Zhang}
\altaffiliation{School of Physics and Astronomy, Sun Yat-sen University, Zhuhai 519082, China}
\altaffiliation{CSST Science Center for the Guangdong-Hong Kong-Macau Greater Bay Area, Zhuhai 519082, China}

\author[0000-0002-8337-4117]{Yaqian Wu}
\altaffiliation{Key Laboratory of Optical Astronomy, National Astronomical Observatories, Chinese Academy of Sciences, Beijing 100012, China}

\correspondingauthor{Suk-Jin Yoon}
\email{sjyoon0691@yonsei.ac.kr}

\begin{abstract}

Using Gaia DR3 data and the wavelet transformation technique, we study the substructures of the Hercules moving group (HMG): Hercules 1 (H1) and Hercules 2 (H2).
Spectroscopic survey data from LAMOST, APOGEE, and GALAH are used to obtain metallicities and ages of stars belonging to the HMG.
Our analysis leads to several key findings as follows:
($a$) the HMG is on average richer in metallicity than the Galactic disk, with H2 being metal richer than H1;
($b$) the HMG likely has a radial metallicity gradient distinct from that of the disk;
($c$) the HMG is on average older than the disk, with H2 being older than H1;
($d$) the HMG likely has a radial age gradient distinct from that of the disk; and
($e$) the metallicity and age distributions of the HMG depend mainly on the Galactic radius but show no dependence on the azimuthal velocity.
Taken all together, we conclude that the HMG is composed primarily of stars undergoing radial migration.
We suggest that the HMG is associated with a higher-order dynamical resonance of the bar of the Galaxy.

\end{abstract}


\section{INTRODUCTION}\label{sec1}

The Hercules moving group (HMG) was first identified and studied by \citet{1958MNRAS.118..154E}.
Member stars of moving groups were suspected to originate from disrupted stellar clusters \citep{1958MNRAS.118...65E}.
However, subsequent theoretical models proposed that these moving groups may result from global dynamical mechanisms related to the non-axisymmetry features of the Galaxy, including the bar \citep{1999ApJ...524L..35D, 2000AJ....119..800D, 2001A&A...373..511F, 2007ApJ...664L..31M,2010MNRAS.405..545G, 2017MNRAS.466L.113M,2017ApJ...840L...2P,2018Natur.561..360A, 2019MNRAS.488.3324F, 2019A&A...626A..41M} and spiral arms \citep{2004MNRAS.350..627D, 2005A&A...430..165F, 2005AJ....130..576Q,2006MNRAS.368..623M,2009ApJ...700L..78A,2011MNRAS.418.1423A,2011MNRAS.414.1607L, 2018MNRAS.481.3794H, 2018ApJ...863L..37M, 2018MNRAS.480.3132Q, 2019MNRAS.484.4540H,2020ApJ...888...75B} or from a combination of the bar and spiral arms \citep{2007A&A...467..145C,2008A&A...488..161C,2018MNRAS.477.5612W,2019MNRAS.485L.104M}.
There are also scenarios suggesting that they are related to the phase mixing caused by the external perturbation
\citep{2018Natur.561..360A, 2018MNRAS.481.3794H, 2018A&A...619A..72R, 2019MNRAS.489.4962K, 2019MNRAS.485.3134L, 2020A&A...643L...3L}.

\citet{2007ApJ...655L..89B} suggested that the HMG consists of a mixture of thin and thick disk stars from the perspective of stellar ages and elemental abundances.
\citet{2018MNRAS.478..228Q}, using GALAH data, suggested that metal-rich stars exhibit the most prominent HMG signature.
\citet{2019MNRAS.484.4540H} supported this conclusion with RAVE data.
\citet{2019A&A...629L...6T} analyzed the Gaia white dwarf population and suggested that the HMG has an age distribution that peaks at 4 Gyr and extends to very old ages.

With Gaia DR2 data \citep{2018A&A...616A..11G}, \citet{2018A&A...619A..72R} found that the HMG has two substructures (Arc8 and Arc9 in their paper) and suggested that the group is related to the outer Lindblad resonance of the Galactic bar due to its approximately conserved vertical angular momentum with radius \citep{2010MNRAS.409..145S,2018MNRAS.478..228Q}.
The purpose of this study is to explore the underlying physical mechanisms responsible for the formation of the two substructures of the HMG within the thin disk of the Galaxy.
In Section \ref{sec2}, we describe the data used for selecting member stars, as well as the data used for determining their metallicity and age.
In Section \ref{sec3}, we present metallicity and age distributions for each of the two substructures.
In Section \ref{sec4}, we discuss various theories in the literature that are relevant to our findings.
Finally, in Section \ref{sec5}, we summarize our new findings and their implications.

\begin{figure}
\includegraphics[width=0.48\textwidth]{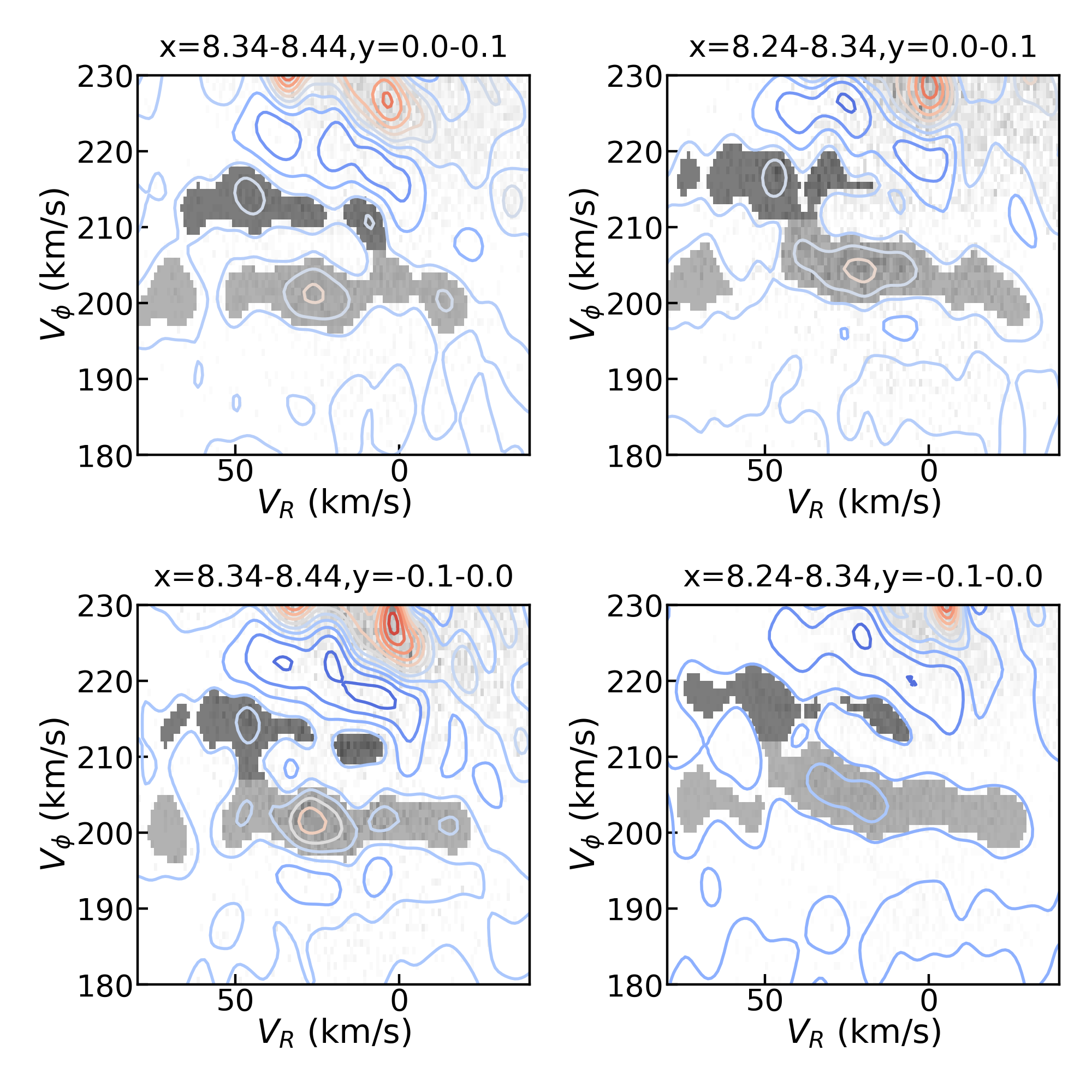}
\caption{The $V_{R}$ versus $V_{\phi}$ distributions of stars in spatial bins. Each subplot represents a spatial bin, as annotated by the subtitle on the top of each subplot. Within each spatial bin, the wavelet transformation is applied to reveal density distributions of stars in the $V_R - V_{\phi}$ coordinate, as shown by the contours. The background 2D histogram shows direct grid distributions of stars, and the darker grayscale indicates higher star counts in the velocity bin. The shaded regions show the selected positions of H1 (relatively darker) and H2 (lighter) according to the wavelet contours.}
\label{xy}
\end{figure}

\section{DATA} \label{sec2}

\begin{figure*}
\plotone{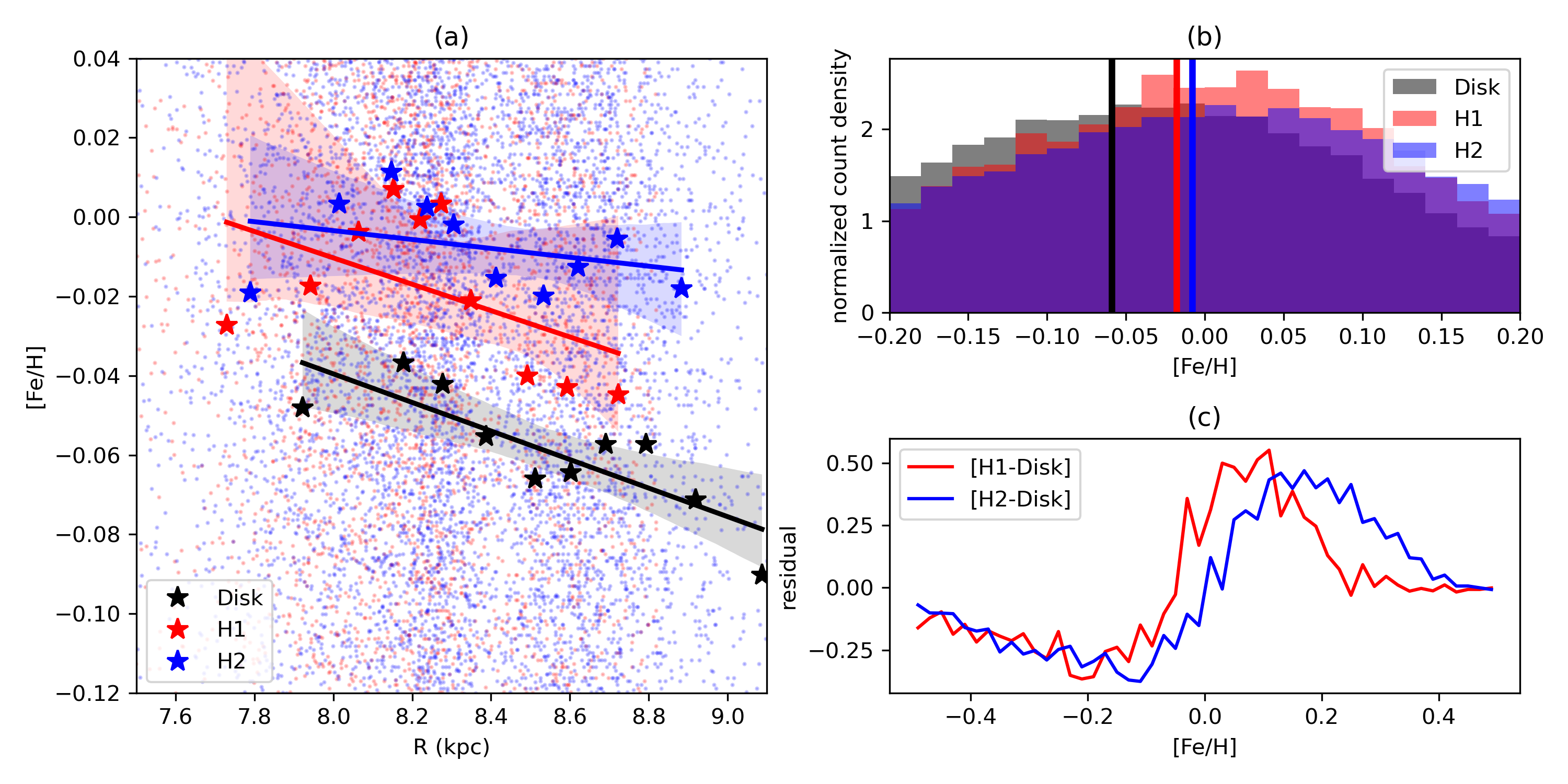}
\caption{
The radial metallicity distributions of member stars of H1 and H2.
Red, blue, and black colors represent H1, H2, and the disk sample, respectively.
(a) The background small dots in the background are the member stars, while star symbols represent median values of $R$ and [Fe/H] in each radial bin.
The 10 radial bins with the same number of stars are used.
Three straight lines and shaded stripes are linear fits of star symbols and their 1$\sigma$ uncertainties, respectively.
(b) We show normalized count density distributions of H1, H2, and the disk sample as functions of [Fe/H]. Three vertical lines show the median metallicity position of each sample.
(c) We show the residuals of [H1 $-$ disk] and [H2 $-$ disk].
}
\label{fehr}
\end{figure*}

\subsection{Position and Velocity}

We select nearby stars from the Gaia DR3 catalog \citep{2016A&A...595A...1G, 2023A&A...674A...1G} with $parallax\_over\_error \geq 5$ and $ruwe<1.4$.
The parallax, proper motion, and radial velocity are used to obtain positions and velocities with a Python package, Astropy \citep{2022ApJ...935..167A}.
The nearby stars within 1.5 kpc with $ \mid z \mid < 200 $ pc and $ \mid V_z \mid < 60 ~\textrm{km s}^{-1}$ are selected as a disk sample.
The $x$, $y$, and $z$ coordinates represent the star's spatial position in the Galactic Cartesian coordinate system, while $V_R$, $V_{\phi}$, and $V_z$ represent the star's velocity components in the Galactic cylindrical coordinate system.

The velocity distribution of stars in each spatial bin is presented in Figures \ref{xy} and \ref{axy}, where wavelet transformation is applied to stars' velocity distribution in the $V_R - V_{\phi}$ coordinate \citep{2018A&A...619A..72R,2023MNRAS.519..432L}.
The disk sample is divided into small bins in the $x$-$y$ plane, with the layout of subplots showing the spatial distribution of the bins.
Figure \ref{axy} in the Appendix displays all the bins, while Figure \ref{xy} shows the bins close to the Sun as examples.

As found by \citet{2018A&A...619A..72R} with the Gaia DR2 data, the HMG has two substructures.
For convenience, we call the one with relatively larger $V_{\phi}$ as Hercules 1 (H1), and the other one as Hercules 2 (H2).
In Figure \ref{xy} and Figure \ref{axy} in the Appendix, the subtitle on the top of each subplot indicates the spatial range in the unit of kiloparsecs that the binned sample covers.
Square bins with a width of 100 pc are used for stars close to the Sun, while larger bin sizes are used for stars far away from the Sun to better reveal the substructures of the HMG.
We note that the HMG has a radially decreasing $V_{\phi}$ distribution, and thus the large radial-bin size can cause a mixing effect.
Therefore, only those bins obviously containing H1 are taken into consideration, and H2 can distribute larger spatial regions than these bins.

Each subplot in Figure \ref{xy} and Figure \ref{axy} has three layers. The bottom layer displays the grid distribution with grayscale representing the number of stars in each small velocity bin.
We use the grid function from the Numpy package \citep{2020Natur.585..357H} to divide stars in the $V_R - V_{\phi}$ coordinate into small velocity bins with a bin size of $1 ~\textrm{km s}^{-1}$.
In each velocity bin, the darker color corresponds to a higher number of stars, while velocity bins with less than two stars are left blank.
The middle layer shows the contour of the wavelet transformation coefficient, with the redder (bluer) contours indicating the larger (smaller) wavelet coefficients.
At the top layer, the shaded region shows selected positions of H1 (relatively darker) and H2 (lighter).
The wavelet transformation coefficient is used to manually select regions related to H1 and H2, and then the stars in shaded regions are selected as potential members of these substructures.
Although there is an artificial interference during the process of selecting member stars, it is assumed that the selected stars are sufficient to statistically represent the HMG.
The goal of this study is not to select all member stars of H1 and H2 but to statistically reveal their metallicity and age properties.

Overall, H1 and H2 distribute farther in the direction toward the Galactic center than in the outward direction.
Moreover, there are more member stars in spatial bins on the negative $y$ side than ones on the positive $y$ side.
This observation supports the theory that the velocity structure, including both H1 and H2, is made of stars vibrating at the Lagrangian points of the Galactic bar \citep{2020ApJ...890..117D}, which is located on the negative $y$ and smaller radius side of the Sun.
H1 has a smaller spatial distribution than H2, but both become more prominent toward the Galactic center. This suggests that if the HMG originated from a resonance, its resonance radius is smaller than the solar radius.

\subsection{Metallicity and Age}

The metallicity and age information allows for a more comprehensive analysis of the properties of H1 and H2.
The wavelet transformation is used to select possible member stars of H1 and H2, and a total of 118,302 and 246,140 member stars are selected for H1 and H2, respectively. These member stars are then cross-matched with APOGEE DR17 \citep{2022ApJS..259...35A}, LAMOST DR9 \citep{2012RAA....12.1197C, 2012RAA....12..723Z, 2015RAA....15.1095L}, and GALAH DR3 \citep{2021MNRAS.506..150B} to obtain metallicity.
We use the Topcat software \citep{tay05} to cross-match different catalogs.
As is well known, there are some systematic differences between these three data sets, and adjustments are made to ensure consistency.
The metallicity difference between APOGEE DR17 and LAMOST DR9 is smaller than the differences between GALAH DR3 and the other two data sets.
Thus, first, we fit the metallicity difference between APOGEE DR17 and LAMOST DR9 against metallicity from the LAMOST DR9 data set with a linear function.
There are 139,904 common stars with [Fe/H] between $-0.8$ and 0.5 dex. The fitted linear function is $\Delta[\rm{Fe/H}]= -0.108\times [\rm{Fe/H}]_{lamost} - 0.022$, after removing stars outside of 3$\sigma$.
Then we modify metallicities from LAMOST DR9 to be consistent with APOGEE DR17.
Second, we fit the metallicity difference between GALAH DR3 and the other two data sets against metallicity from the GALAH DR3 data set with a linear function.
There are 73,891 common stars with [Fe/H] between $-0.8$ and 0.5 dex.
The fitted linear function is $\Delta[\rm{Fe/H}]= -0.152 \times [\rm{Fe/H}]_{galah} - 0.008$, after removing stars outside of 3$\sigma$.
Then we modify the metallicity from GALAH DR3 to be consistent with the other two data sets.
Finally, we obtain 9\,371 and 19\,891 stars of H1 and H2 with metallicity information, respectively.
On the other hand, we cross-match the member stars of H1 and H2 with the age samples from \citet{2021MNRAS.501.4917W} for giant stars and from \citet{2019ApJ...887...84Z} for dwarf stars.
We obtain age information for 209 giant stars and 282 dwarf stars for H1 and 476 giant stars and 654 dwarf stars for H2.

\section{RESULTS} \label{sec3}

Figure \ref{fehr}(a) shows the metallicity distribution of the member stars of H1 and H2 at different Galactic radii.
Red and blue dots in the background represent member stars for H1 and H2, respectively.
These stars are divided into 10 bins along the $R$ axis, with each bin containing an equal number of stars.
The median values of $R$ and metallicity are taken to represent the stars in the bin, shown as star symbols in red for H1, blue for H2, and black for the disk sample.
To check the adopted number of bins affecting the result, we also try 20 and 30 radial-bin cases as well, which have the same result as the 10 bin case.
Three straight lines are fitted to these star symbols to better illustrate the differences among the H1, H2, and disk samples.
The shaded regions depict 1$\sigma$ uncertainty of the fitted lines, which are obtained through the jackknife analysis.
Both H1 and H2 are metal richer than the disk sample, while H2 is metal richer than H1.
We note that the star symbols exhibit a wavelike pattern of metallicity against Galactic radius.
It appears that the feature is related to the Milky Way spiral arms \citep{2022A&A...663A..38K}.

Figure \ref{fehr}(b) shows normalized number distributions of the H1, H2, and disk samples as functions of [Fe/H].
Three vertical lines indicate the median metallicity positions of the samples.
Although the differences between the median metallicity values are not large, there is a trend for both H1 and H2 to be metal richer than the disk sample and for H2 to be slightly metal richer than H1.
To test whether the three samples are drawn from the same distribution, the Kolmogorov–
Smirnov test is conducted, which gives $p$-values of 1.4e-75, 3.5e-215, and 3.2e-16 for the comparisons of the disk and H1 samples, the disk and H2 samples, and the H1 and H2 samples, respectively. Therefore, these samples have different distributions. If using binned values to represent the total samples as shown in the Figure \ref{fehr}(b), we would get $p$-values of 1.4e-13, 7.8e-27 and 0.0005 for the comparisons of the disk and H1 samples, the disk and H2 samples, and the H1 and H2 samples, respectively. The less bins used, the larger p-values would we get.
Figure \ref{fehr}(c) shows the residual distributions of [H1 $-$ disk] and [H2 $-$ disk].
The H1 and H2 samples are generally metal richer than the disk sample,
and the H2 sample is metal richer than the H1 sample.


\begin{figure}
\includegraphics[width=0.48\textwidth, angle=0]{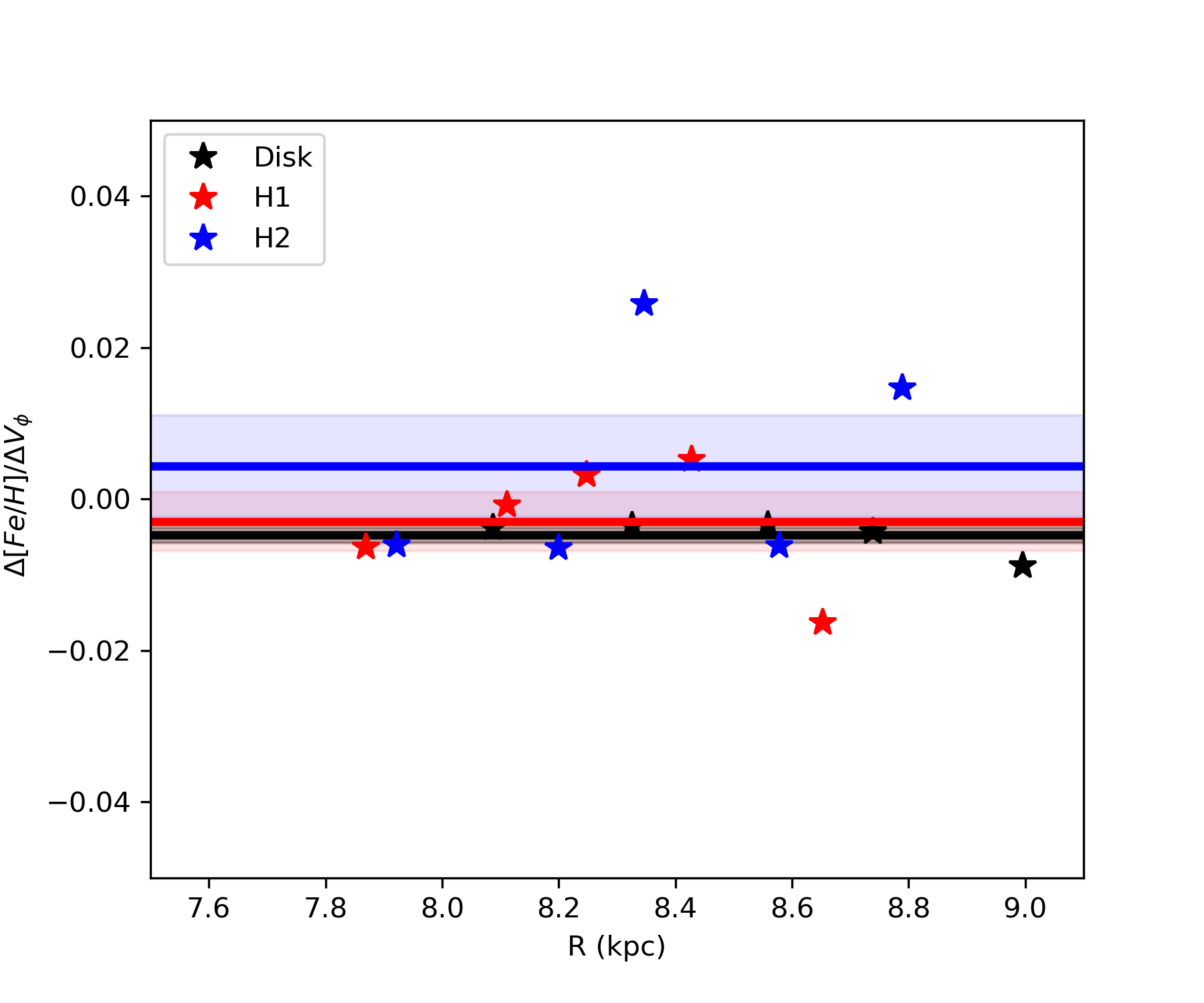}
\caption{
The metallicity gradients against $V_{\phi}$ distributions of member stars of H1 and H2 in radial bins.
Red, blue, and black colors represent H1, H2, and the disk sample, respectively.
The star symbols represent the slopes of fitted straight lines for [Fe/H] versus $V_{\phi}$ in corresponding radial bins.
Three horizontal lines and shaded stripes respectively show the mean values and the 1$\sigma$ uncertainty of the mean value.}
\label{fehvphir}
\end{figure}

\begin{figure*}
\plotone{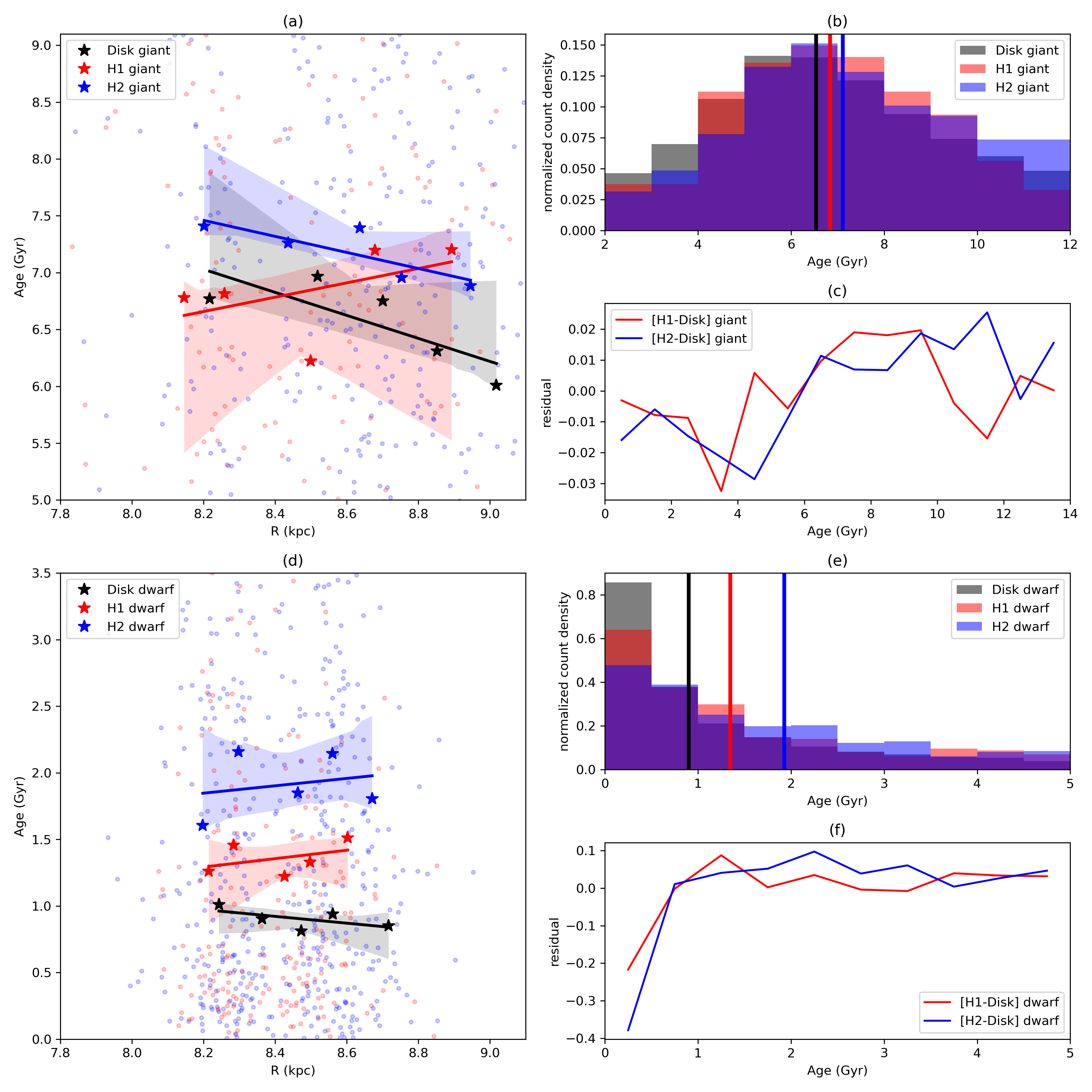}
\caption{The same as Figure \ref{fehr}, but for age.
This figure shows radial age distributions of the member stars of H1 and H2.
Red, blue, and black colors represent H1, H2, and the disk sample, respectively.
In the left two panels (a) and (d), the background small dots are the member stars, while star symbols represent the median values of $R$ and age in each radial bin.
The five radial bins with the same number of stars are used.
Three straight lines and shaded stripes are linear fits of star symbols and their 1$\sigma$ uncertainties, respectively.
Similarly to metallicity distribution, we also show normalized count density distributions and the residuals for age distributions in the right panels (b), (c), (e), and (f).
}
\label{ager}
\end{figure*}

\begin{figure*}
\plotone{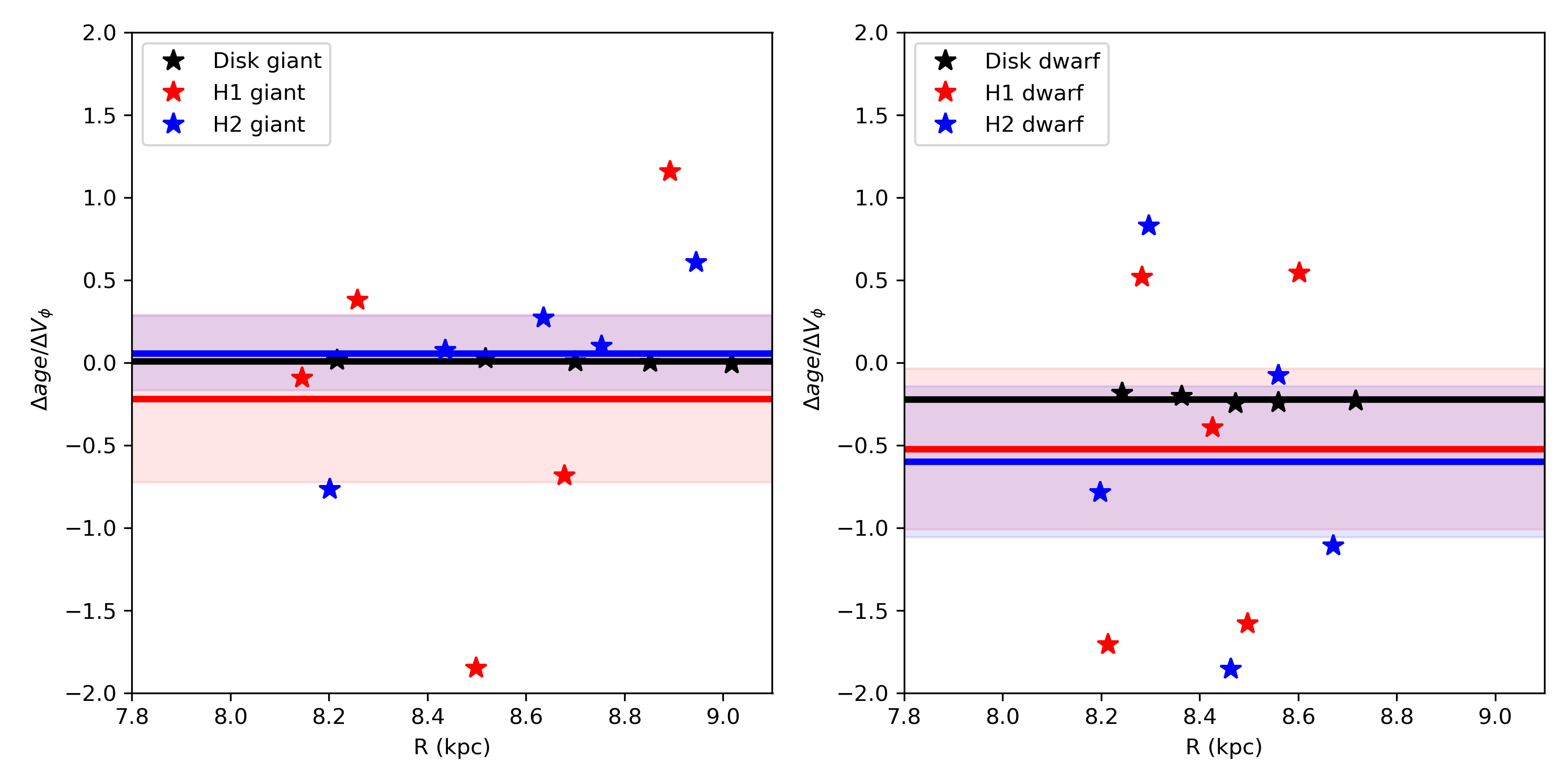}
\caption{
The age gradients against $V_{\phi}$ of the member stars of H1 and H2.
Red, blue, and black colors represent H1, H2, and the disk sample, respectively.
The star symbols represent the slopes of fitted straight lines for age versus $V_{\phi}$ in corresponding radial bins.
Three horizontal lines and shaded stripes respectively show the mean values and the 1$\sigma$ uncertainties.}
\label{agev}
\end{figure*}

The metallicity and the $V_{\phi}$ distributions of the HMG both have negative slopes against the Galactic radius \citep{2014A&A...563A..60A}.
Hence, using all member stars of H1 and H2 for fitting [Fe/H] against $V_{\phi}$ would suffer a mixing effect.
The negative slope against the Galactic radius will cause an overall positive slope against $V_{\phi}$.
In Figure \ref{fehvphir}, we show the slopes (denoted by $\Delta$[Fe/H]/$\Delta V_{\phi}$) of fitted straight lines for [Fe/H] versus $V_{\phi}$ in each radial bin (star symbols).
These slopes are taken as metallicity gradients against $V_{\phi}$ in each radial bin.
Three horizontal lines show the mean values of these star symbols.
If metallicity depends on $V_{\phi}$ within each bin, the mean gradient should be different from zero.
The differences between the mean values of the gradients and zero are smaller than the standard deviation of these slopes, indicating that both H1 and H2 have nearly zero metallicity gradients against $V_{\phi}$.

Overall, Figures \ref{fehr} and \ref{fehvphir} indicate that H1 and H2 are two independent velocity structures with different metallicities, and their metallicity distributions depend dominantly on the Galactic radius rather than on the azimuthal velocity component $V_{\phi}$.
This implies that H1 and H2 are two distinct structures, rather than artificially separated parts of a larger velocity structure of the HMG, in which metallicity continuously becomes poorer with increasing $V_{\phi}$ \citep{2021MNRAS.505.2412C}. The observed metallicity differences between H1 and H2 are likely due to differences not in their kinematics but in their formation histories.

In Figure \ref{ager}, we show the age distributions of member stars in H1 and H2 separately for giant stars and dwarf stars and compare them to the age distribution of the disk sample.
The methods used to obtain age for giant stars and dwarf stars are essentially different, and there is a systematic difference between them.
Figure \ref{ager}(a) shows the ages of member giant stars of H1 and H2 at different Galactic radii.
Similarly to the treatment for metallicity, we divide them into five bins along the $R$ axis with each bin having the same number of stars.
Then the median values of $R$ and age are taken to represent stars in the bin.
To check the adopted number of bins affecting the result, we also try 10 and 20 radial-bin cases and they have the same result as the five bin case.
Red, blue, and black star symbols represent H1, H2, and the disk sample, respectively.
The straight lines are the linear fits of these star symbols to better illustrate the difference among the H1, H2, and disk samples.
The shaded regions show the uncertainty of fitted lines, which is obtained by the jackknife analysis.
The figure shows that the giant stars of H2 are on average older than H1 and the disk stars.
Giant stars of H1 are on average older than the disk stars at a large Galactic radius but younger at a small radius.
Figure \ref{ager}(b), similar to the metallicity distribution in Figure \ref{fehr}(b), shows the normalized number distributions of ages for giant stars in the H1, H2, and disk samples, with vertical lines indicating the median age position of each sample.
Figure \ref{ager}(c) shows the residuals of [H1 $-$ disk] and [H2 $-$ disk].
H2 has a larger proportion of stars at older ages compared to H1 and the disk sample.
The right two panels (b,c) show that both H1 and H2 are generally older than the disk sample and H2 is slightly older than H1.
Figure \ref{ager}(d-f) are the same as Figure \ref{ager}(a-c) but for the dwarf stars.
Overall, although the age information itself has large uncertainty and the age distribution has large dispersion, there is a sign that the differences between the disk sample and two Hercules substructures grow larger with increasing Galactic radius.

Similar to the metallicity distribution, the age distributions also depend on the Galactic radius.
To avoid the mixing effect caused by the dependence of $V_{\phi}$ on the Galactic radius, the slope of age versus $V_{\phi}$ is fitted in each radial bin.
In Figure \ref{agev}, we show the slopes (denoted by $\Delta$age/$\Delta V_{\phi}$) of fitted straight lines for age versus $V_{\phi}$ in each radial bin (star symbols).
These slopes are taken as age gradients against $V_{\phi}$ in corresponding radial bins.
The horizontal lines show the mean values of gradients.
The differences between the mean values of the gradients and zero are much smaller than the standard deviation of these slopes, indicating that both H1 and H2 have nearly zero age gradients against $V_{\phi}$. Overall, our results indicate that H1 and H2 have different ages, and their age distributions depend dominantly on the Galactic radius rather than on the azimuthal velocity component, $V_{\phi}$.

In summary, H1 and H2 have on average higher metallicities and older ages than the disk sample at their respective radii. Hence, we suggest that H1 and H2 are composed of stars that were born in the inner Galaxy and migrated outwards to their current locations.

\section{DISCUSSION} \label{sec4}

First, we introduce some concepts relevant to the dynamical resonance based on the epicycle approximation theory (\citet{2008gady.book.....B}, Chapter 3.3.3).
The guiding center corotates along with the potential when $ \Omega_0 = \Omega_p$, where $\Omega_0$ is the circular frequency at Galactic radius $R$, and $\Omega_p$ is the steady pattern speed at which the potential rotates.
The Lindblad resonance occurs when $ \pm m(\Omega_0 - \Omega_p) = \kappa_0$, where $m$ is an integer and $\kappa_0$ is the epicycle frequency.
The minus sign before $m$ corresponds to the outer Lindblad resonance, and the resonance with $- m$ is usually called the outer $m:1$ resonance ($m:1$ OLR; \citet{2000AJ....119..800D}).

Many studies have used the HMG to obtain parameters for the non-axisymmetric part of the Milky Way potential, such as the bar and the spiral arms.
\citet{2000AJ....119..800D} suggested the bimodality of velocity distribution (i.e., the Hercules velocity structure and the central low-velocity part) is caused by the outer 1:1 Lindblad resonance of the bar, and the bar's derived pattern speed was $53 ~\textrm{km s}^{-1} \textrm{kpc}^{-1}$.
\citet{2007ApJ...664L..31M} reported that the Hercules velocity structure is caused by 2:1 OLR of the bar and estimated the bar's pattern speed to be $\Omega_b/\Omega_0 = 1.87$, similar to that obtained by \citet{2000AJ....119..800D}.
Such a fast pattern speed of a short bar was consistent with the observed length of the bar at that time.
\citet{2010MNRAS.405..545G} investigated the dynamical effect using the two-bar model with one short bar plus the other longer, flatter bar and obtained pattern speeds of 55.9 and $54.9 ~\textrm{km s}^{-1} \textrm{kpc}^{-1}$ for the traditional bar and long bar, respectively.
\citet{2017ApJ...840L...2P} suggested that the Hercules velocity structure is caused by the corotation resonance of the bar with a pattern speed of $39 ~\textrm{km s}^{-1} \textrm{kpc}^{-1}$.
This is consistent with a long and slowly rotating bar observed by recent photometric and spectroscopic surveys of the inner Galaxy.
Later, the long, slow bar model was favored by many simulations.
\citet{2019A&A...632A.107M} showed that the Hercules velocity structure could be associated with orbits trapped at the corotation resonance of the bar with a pattern speed of $39 ~\textrm{km s}^{-1} \textrm{kpc}^{-1}$.
\citet{2020MNRAS.495..895B} favored trapping at the corotation of the bar with a pattern speed of $36 ~\textrm{km s}^{-1} \textrm{kpc}^{-1}$.
\citet{2020ApJ...890..117D} proposed that the Hercules is made of Trojans, stars captured by the Lagrangian points of the stellar bar with a pattern speed of $40 ~\textrm{km s}^{-1} \textrm{kpc}^{-1}$.
\citet{2005AJ....130..576Q} suggested that the Hercules velocity structure is caused by the inner 4:1 Lindblad resonance (4:1 ILR) of spiral arms and estimated the pattern speed for spiral arms as $\Omega_s/\Omega_0 = 0.66$, namely $18.1 ~\textrm{km s}^{-1} \textrm{kpc}^{-1}$.
\citet{2018A&A...615A..10M} studied the combined dynamical effects of the bar and the spiral arms.
In their simulations, the same pattern speed of $28.5 ~\textrm{km s}^{-1} \textrm{kpc}^{-1}$ was assumed for the bar and the spiral arms.
Their result suggests that the HMG is associated with 8:1 ILR of the spiral arms, rather than the bar and bulge perturbations.

Overall, these past results showed that the HMG can be used to obtain parameters like the pattern speed for the bar or the spiral arms.
However, the simulations did not consider the substructures of the HMG due mainly to observational limitations in their time.

After two substructures of the HMG were observed by \citet{2018A&A...619A..72R}, new scenarios have been put forward to explain them. \citet{2018ApJ...863L..37M} suggested that the Hercules substructures, including H1, H2, and HR1614 moving groups, are formed mainly by the 12:1, 8:1, and 6:1 ILRs of the spiral arms \citep{2020ApJ...888...75B}.
According to their theory, even the Local Arm originates from the corotation resonance of the main spiral arms \citep{2017ApJ...843...48L}.
It is natural that the kinematic structures result from stellar orbits trapped by the strong spiral arms' resonances.
Their results showed that the Coma Berenices, Pleiades, and Hyades moving groups are associated with the corotation resonance, while the HMG is created by the bulk of high-order ILRs.
They also suggested that the age of the corotation zone could be about 3 Gyr and the corotation region is close to a closed-box system, in which the metallicity increases rapidly with time.
Our results that the metallicity of older H2 is richer than that of the younger H1 do not support this prediction.
Our stars of H1 and H2 do not show an age enhancement within 3 Gyr.

On the other hand, simulations considering high-order resonances of the bar can also generate substructures of the HMG \citep{2019A&A...626A..41M,2020MNRAS.499.2416A}.
The simulations suggested that H1 and H2 are composed of stars trapped in the 4:1 and 5:1 OLR, which favors a slow pattern speed.
\citet{2021MNRAS.500.4710C} investigated resonance sweeping of a decelerating bar using test particle simulations based on the secular perturbation theory \citep{2021MNRAS.505.2412C}.
They obtained the bar’s current pattern speed of $\Omega_b = 35.5 ~\textrm{km s}^{-1} \textrm{kpc}^{-1}$ with a slowing rate $\dot{\Omega}_b = -4.5 ~\textrm{km s}^{-1} \textrm{kpc}^{-1}\textrm{Gyr}^{-1}$.
When the bar decelerates, the resonances migrate outward \citep{2018A&A...616A..86H,2021MNRAS.500.4710C}.
The resonance sequentially captures surrounding stars and traps them, forming a so-called tree-ring structure \citep{2021MNRAS.505.2412C}.
This model also predicted the age and metallicity distributions in the $V_R - V_{\phi}$ coordinate \citep{2017A&A...601A..59A, 2021MNRAS.505.2412C}, which is roughly consistent with our observed result.
However, the tree-ring theory does not explain why the HMG has two substructures.

\citet{2022A&A...667A.116B} suggested that moving groups can be thought of as continuous manifolds in the 6D phase space.
The ridges in the $R - V_{\phi}$ coordinate and arches in the $V_R - V_{\phi}$ coordinate are regarded as specific projections of these manifolds.
They suggested that stellar velocity distribution in the solar neighborhood is possibly a sign of incomplete vertical phase-mixing processes of previous accretion events.
This model did not predict the different ages of H1 and H2, which is at odds with our results.
\citet{2020ApJ...890...85L} found a clear $z - V_z$ phase-space snail-shell feature only in H1 (not in H2) and suggested that the two substructures may have experienced different formation processes.
\citet{2023A&A...673A.115A} suggested that multiple perturbations at different times or from different perturbers could have affected the phase space.
\citet{2023MNRAS.521..114T} suggested the dynamical effect of small disturbances can be erased by scattering on a gigayear time-scale.
\citet{2023MNRAS.521L..24D} suggested that these fragile structures in the phase space may be destroyed by a rotating bar.

In this study, we propose that H1 and H2 are related to the radial migration \citep{2022ApJ...936L...7C}.
Radial migration can be associated with the spiral waves \citep{2002MNRAS.336..785S}, resonant scattering with transient spiral arms \citep{2008ApJ...684L..79R}, the resonance overlap of the bar and the spiral arms \citep{2010ApJ...722..112M,2011A&A...527A.147M}, and/or the corotation resonance of the bar \citep{2012A&A...548A.126M}.
\citet{2020A&A...638A.144K}, using simulations, studied the effect of the slowing down of a stellar bar migrating from the inner to the outer disk.
They suggested that the slowing-down bar enables stellar particles trapped in the main resonances (corotation and OLR) to propagate outward across the disk.

The radial migration theory can explain the observed metallicity and age distributions of H1 and H2.
According to this theory, metal-rich stars that migrate outward from the inner radius would make H1 and H2 metal richer than the disk sample. This is because stars that have migrated from the inner regions of the Galaxy should have richer metallicity due to the radial metallicity gradient of the disk.
Hence we suggest that H2 has been part of radial migration for a longer period of time since it apparently has a smaller radial metallicity gradient than the disk sample. This implies that H2 has been located farther from the inner radius of the galaxy for a longer time, allowing it to accumulate more metal-rich stars by radial migration over an extended period.
On the other hand, H1 is suggested to have started its migration recently as it has a radial metallicity gradient only slightly smaller than that of the disk.
This suggests that H1 has migrated from the inner regions of the Galaxy more recently compared to H2, and therefore has accumulated less metal-rich stars by radial migration.
Furthermore, older stars have had more time for radial migration and thus would have migrated farther from their original birthplaces compared to younger stars. This would make the stars of H1 and H2, located at larger radii, older than their disk counterparts, which is consistent with the observed age distributions of these substructures.
Overall, the proposed radial migration theory provides a possible explanation for the observed metallicity and age distributions of H1 and H2 in the Milky Way's disk.

\section{CONCLUSION} \label{sec5}

We have investigated the metallicity and age distribution of H1 and H2.
Our findings are summarized as follows.
\begin{itemize}
    \item H1 has a smaller spatial range compared to H2. Both H1 and H2 distribute farther inwardly than outwardly, and their wavelet transformation features are stronger in the negative $y$ bins than in their corresponding positive $y$ bins.
    \item H2 appears to have a richer and flatter radial metallicity distribution than H1, while H1 appears to have a richer and flatter radial metallicity distribution than the disk sample. This suggests that H2 may have accumulated more metal-rich stars through radial migration over time, while H1 has accumulated fewer metal-rich stars.
    \item There is no clear tendency of metallicity gradient against $V_{\phi}$ within H1 and H2. This implies that H1 and H2 are two distinct structures with different metallicities, rather than being part of a larger velocity structure of the HMG, in which the metallicity is expected to continuously decrease as $V_{\phi}$ increases.
    \item H2 is on average older than H1 and the disk sample. This suggests that H2 stars formed earlier and had more time to evolve and migrate radially compared to H1 stars and the disk sample.
    \item Stars in both H1 and H2 are likely to have radial age gradients different from the disk stars. This implies that the age distributions of H1 and H2 stars are distinct from the overall disk population,
    \item There is no clear tendency of the age gradient against $V_{\phi}$ within H1 and H2. This further supports the idea that they are dynamically separate structures.
\end{itemize}

Based on these findings, we suggest that H1 and H2 are most likely two distinct velocity structures and that higher order of resonance should be considered regardless of their association with the bar and/or the spiral arms.
It is proposed that H1 and H2 are mainly composed of stars undergoing radial migration, which originate from the corotation region of the bar.
The metallicity and age distributions of H1 and H2 are roughly consistent with the pattern speed of potentially slowing down, but more detailed simulations should be done to accord with the observational properties of substructures of the HMG.

\acknowledgments
S.-J.Y. and X.-L.L. acknowledge support from the Mid-career Researcher Program (2019R1A2C3006242) and the Basic Science Research Program (2022R1A6A1A03053472) through the National Research Foundation (NRF) of Korea.

This work has made use of data from the European Space Agency (ESA) mission
{Gaia} (\url{https://www.cosmos.esa.int/gaia}), processed by the {Gaia}
Data Processing and Analysis Consortium (DPAC,
\url{https://www.cosmos.esa.int/web/gaia/dpac/consortium}). Funding for the DPAC
has been provided by national institutions, in particular, the institutions
participating in the {Gaia} Multilateral Agreement.

\appendix

Figure \ref{axy} shows the spatial layout of adopted bins.
The red circles on the $X-Y$ coordinates at the corner of the four panels show their relative spatial positions.
Each subplot shows the $V_R$ versus $V_{\phi}$ distribution of stars in each spatial bin.
The subtitle on the top of each subplot annotates the spatial range in units of kiloparsecs that the bin sample covers.
The position of the Sun is at $x=8.34$ kpc and $y=0$ kpc.
The larger bin size is used for bins far away from the Sun.
For stars in each bin, wavelet transformation is applied to stars' velocity distribution in the $V_R - V_{\phi}$ coordinate.
Each subplot has three layers.
The bottom layer shows grid distribution.
The darker color means more stars in the velocity bin, and the velocity bins with less than two stars are left blank.
The middle layer shows the contour of the wavelet transformation coefficient.
The redder (bluer) contour represents the larger (smaller) wavelet coefficient.
At the top layer, the shaded region shows the selected positions of H1 (relatively darker) and H2 (lighter) in the $V_R - V_{\phi}$ coordinate.
Comparing positions of H1 and H2 in horizontal panels with the same $y$ coordinates, H1 and H2 have relatively larger $V_{\phi}$ distributions in panels with smaller $x$ coordinates.


\counterwithin{figure}{section}

\renewcommand{\thefigure}{A\arabic{figure}}
\setcounter{figure}{0}

\begin{figure}
\centering
\subfigure{
\plotone{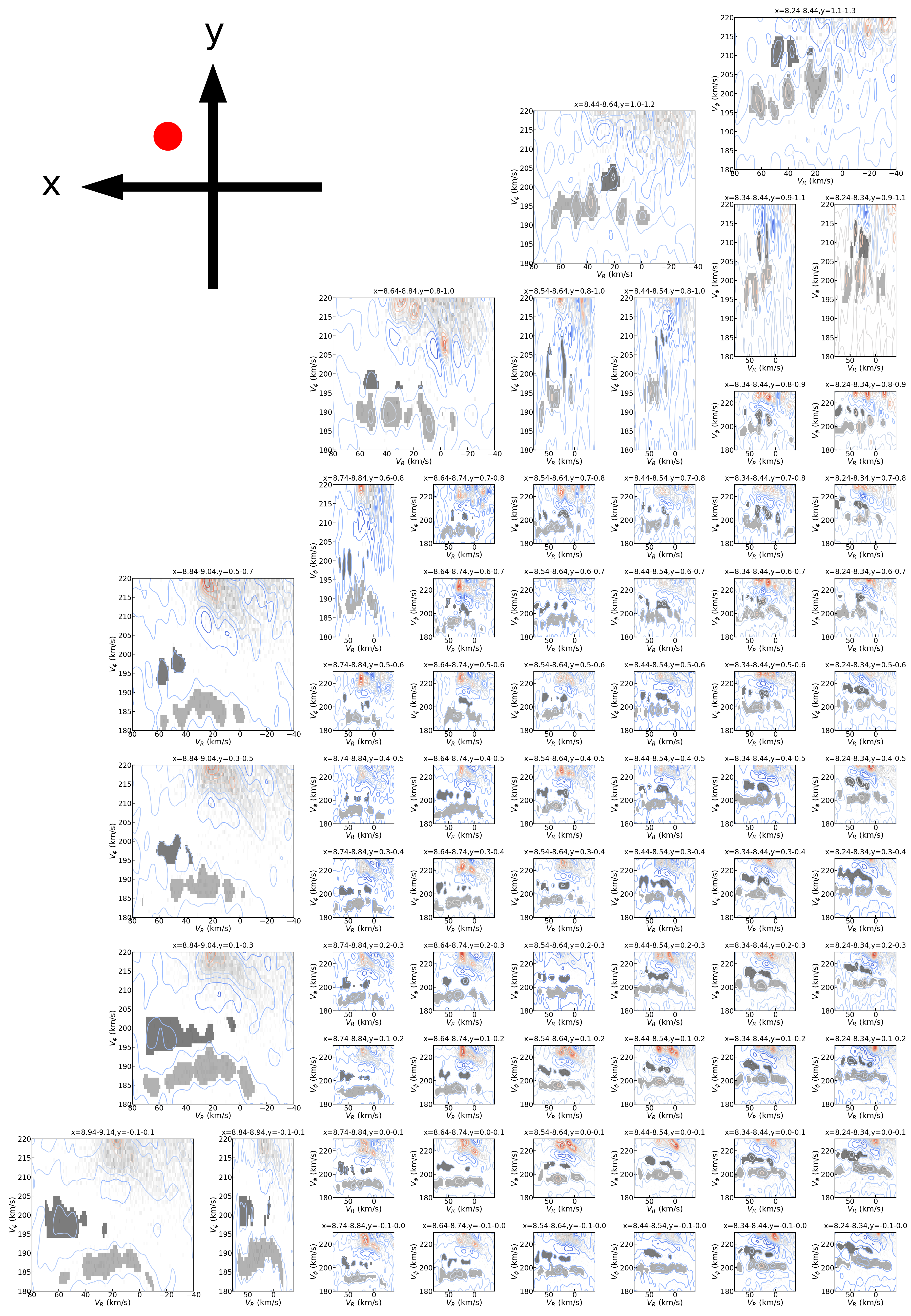}
}
\end{figure}
\begin{figure}
\subfigure{
\plotone{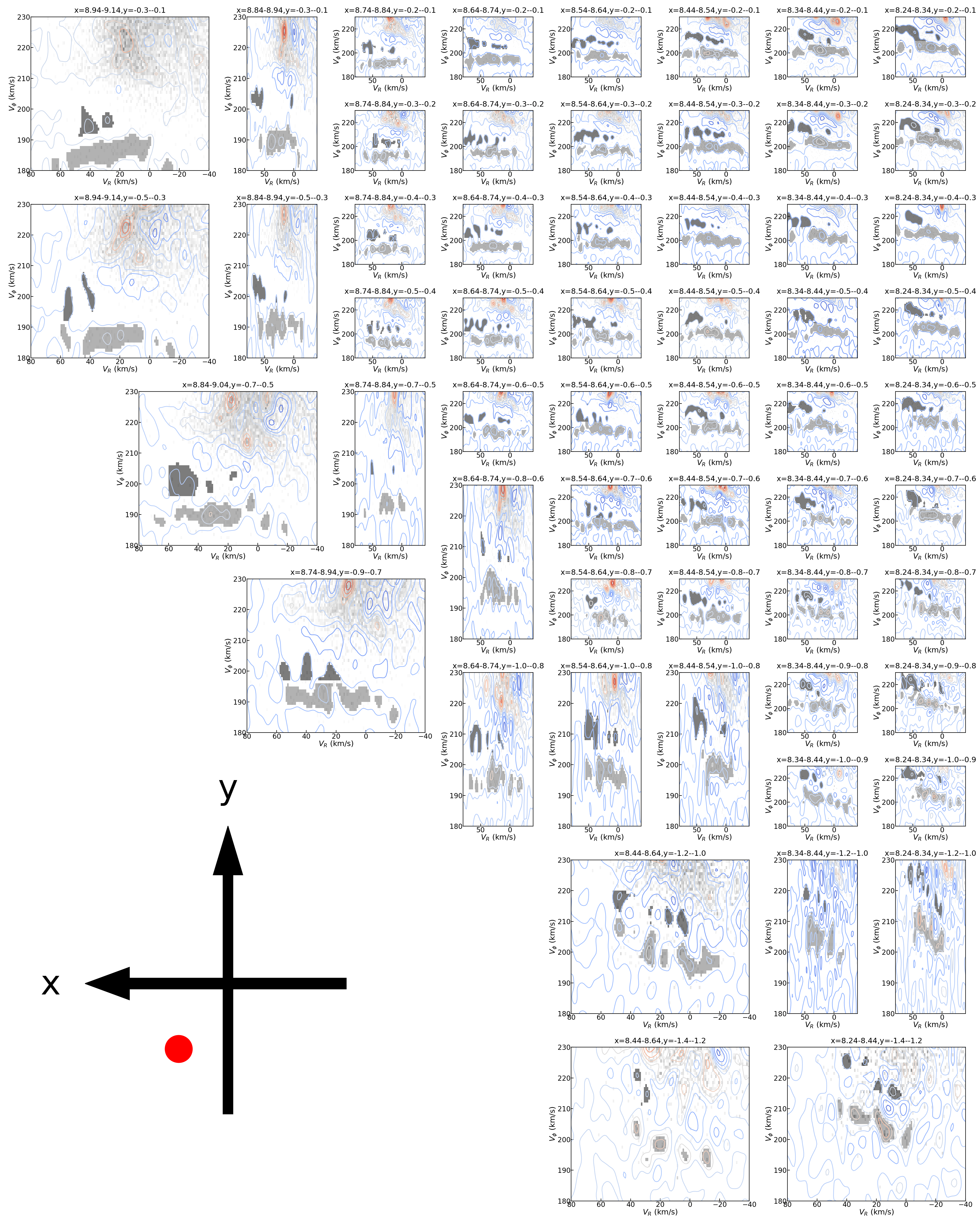}
}
\end{figure}
\begin{figure}
\subfigure{
\plotone{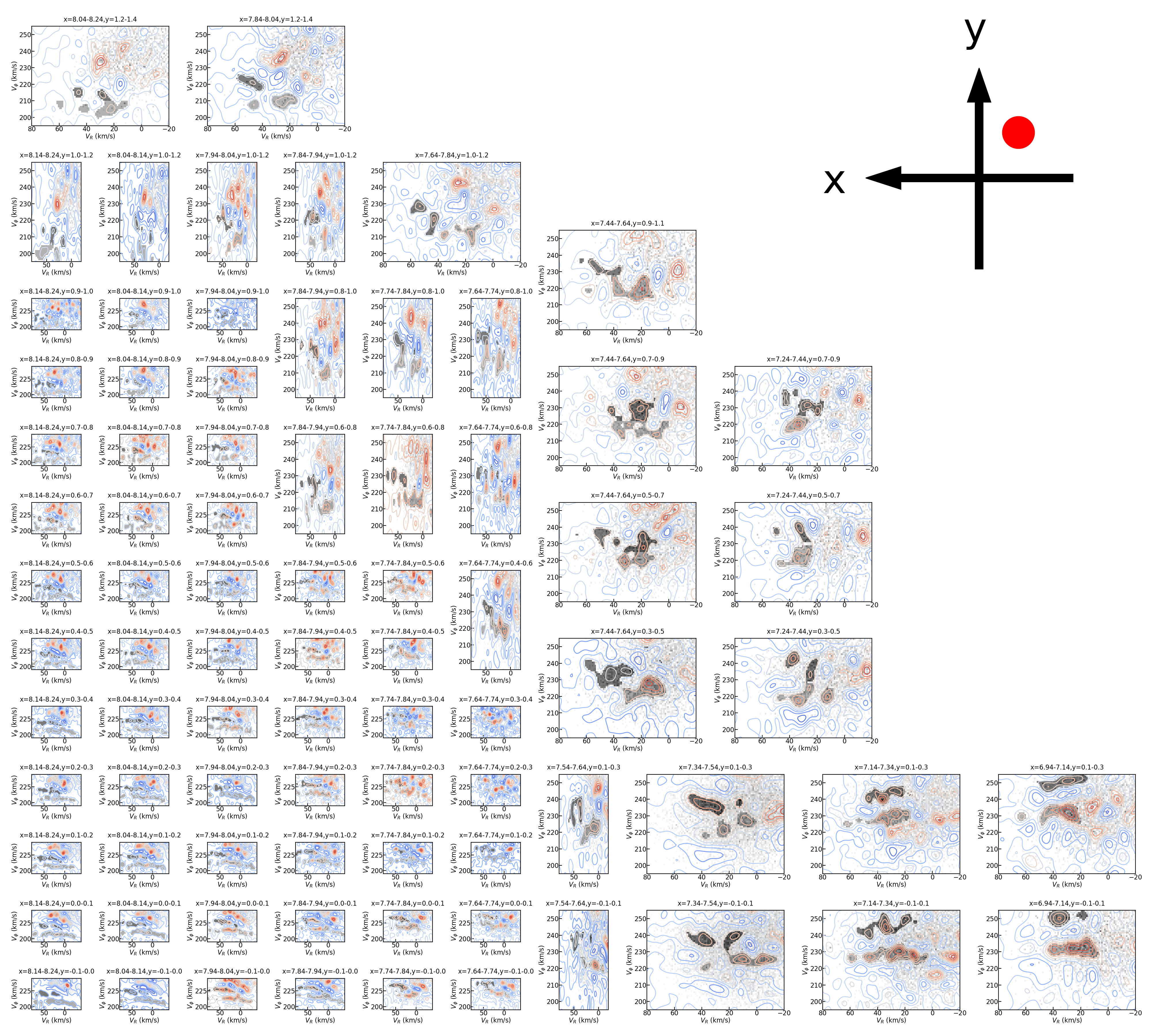}
}
\end{figure}
\begin{figure}
\subfigure{
\plotone{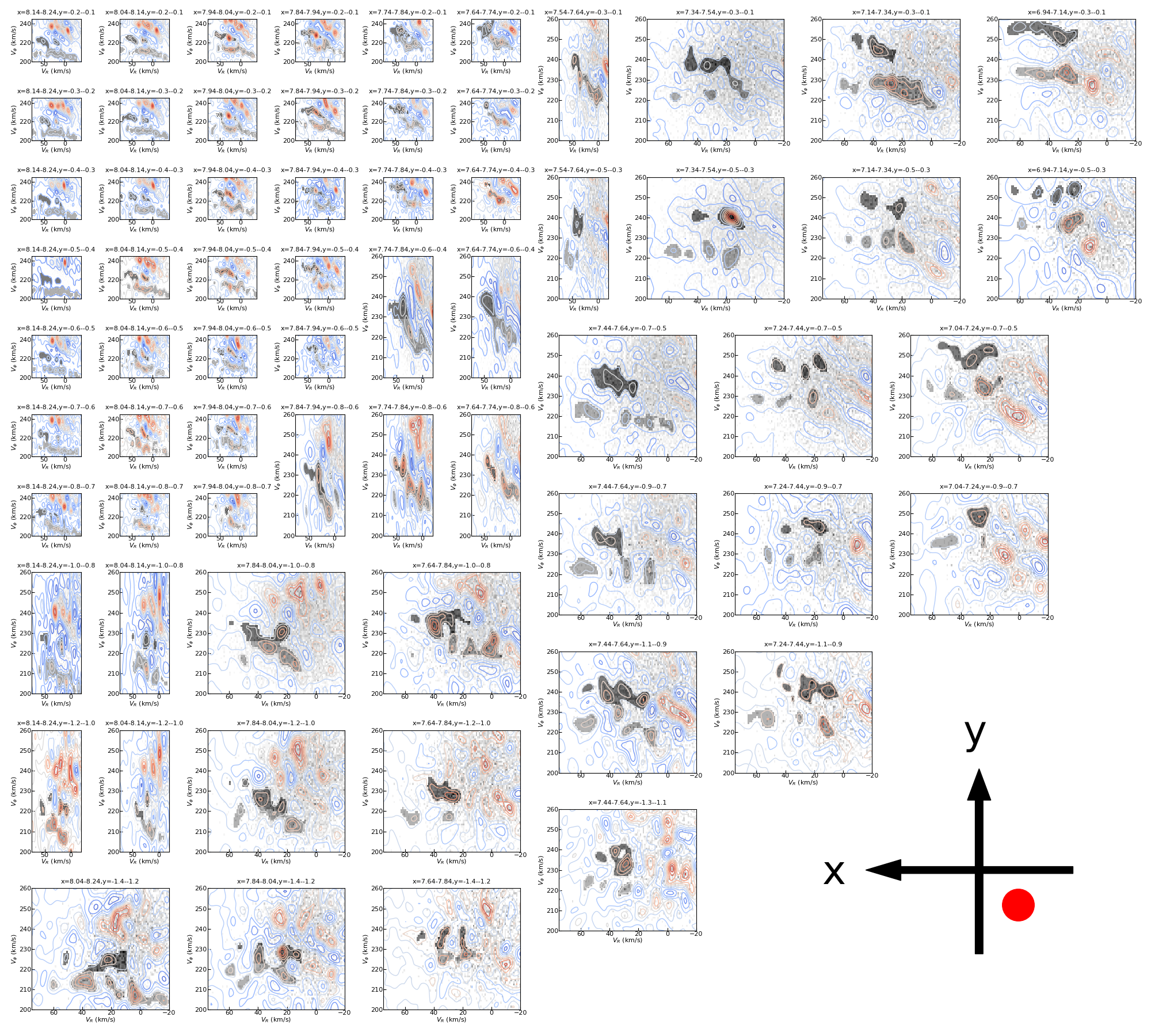}
}
\caption{Velocity distributions of stars in spatial bins. These subplots are similar to those in Figure \ref{xy}, but for all spatial bins. The layout of subplots shows the spatial distribution of bins.}
\label{axy}
\end{figure}

\end{document}